\documentclass[a4paper]{article}

\usepackage{cite,amsmath,amssymb,epsfig}

\newcommand{\beqs}{\begin{equation*}}
\newcommand{\beq}{\begin{equation}}

\newcommand{\eeqs}{\end{equation*}}
\newcommand{\eeq}{\end{equation}}

\newcommand{\beqas}{\begin{eqnarray*}}
\newcommand{\beqa}{\begin{eqnarray}}

\newcommand{\eeqas}{\end{eqnarray*}}
\newcommand{\eeqa}{\end{eqnarray}}

\newcommand{\seq}[5]{ 
\parbox{#1}{\begin{eqnarray*} #2 \end{eqnarray*}} \hfill
\parbox{#3}{\begin{eqnarray*} #4 \end{eqnarray*}} \hfill
\parbox{1cm}{\begin{equation} \label{#5} \end{equation}}}

\newcommand{\eq}[2]{\begin{equation} #1 \label{#2} \end{equation}}

\newcommand{\eqa}[2]{\begin{eqnarray} #1 \label{#2} \end{eqnarray}}

\newcommand{\eps}{\varepsilon}
\newcommand{\al}{\alpha}
\newcommand{\be}{\beta}

\newcommand{\de}{\delta}
\newcommand{\om}{\omega}
\newcommand{\ka}{\kappa}

\newcommand{\Om}{\Omega}

\newcommand{\blist}{\begin{itemize}}

\newcommand{\elist}{\end{itemize}}

\begin{document}



\renewcommand{\thefootnote}{\fnsymbol{footnote}}

\title{${S}$-matrix for ${s}$-wave gravitational 
scattering}
\author{P. Fischer\footnotemark[1], D. Grumiller\footnotemark[1], W. Kummer\footnotemark[1], D.V. Vassilevich\footnotemark[2]}
\footnotetext[1]{Institut f\"ur Theoretische Physik, TU Wien, 
                 Wiedner Hauptstr.  8--10, A-1040 Wien, Austria}
\footnotetext[2]{Institut f\"ur Theoretische Physik, Universit\"at Leipzig,
                 Augustusplatz 10, D-04109 Leipzig, Germany}
\date{\today}
\maketitle

\begin{abstract}
In the $s$-wave approximation the 4D Einstein gravity with
scalar fields can be reduced to an effective 2D dilaton gravity
coupled nonminimally to the matter fields. We study the leading
order (tree level) vertices. The 4-particle matrix element
is calculated explicitly. It is interpreted as scattering
with formation of a virtual black hole state. As one novel feature we 
predict the gravitational decay of $s$-waves.
\end{abstract}



\section{Introduction}

Quantum gravity is beset with well-known conceptual problems. Probably the most
challenging one is the dual r{\^o}le of geometric variables as fields which at 
the same time determine the local and global properties of the manifold on 
which they act. Due to this fact and because gravity is perturbatively 
non-renormalizable it is desirable to use non-perturbative methods. 
Unfortunately, in $d=4$ this is technically problematic. 

Therefore, models in $d=2$ are considered frequently in this context, most of 
which lack an important feature present in ordinary gravity: They contain no 
continuous physical degrees of freedom. One way to overcome this without 
leaving the comfortable realm of two dimensions is the inclusion of matter. 

The aim of the present work is to shed some light on a 2d system which is
closely related to Einstein gravity in $d=4$ and thus of some phenomenological 
relevance, namely the spherically reduced Einstein-massless-Klein-Gordon 
model. It exhibits many interesting properties already at the classical level.
Our exact treatment of the geometric part allows for the straightforward 
calculation of the non-local vertex, which is interpreted as the exchange of a 
virtual black hole. The (highly non-trivial) classical $S$-matrix resulting 
from this graph is determined and discussed in this paper. Furthermore our 
approach provides the basis for quantum corrections in the matter sector.

\section{First order formulation}

When the line element\footnote{$x := (x^0,x^1)$ and the indices $\al,\be$ go from 0 to 1}
\eq{
(ds)^{2}_{(4)}=g_{\al \be }(x)dx^{\al }dx^{\be }-X(x)d\Om ^{2}_{S^{2}}
}{le}
only depends on the metric on the unit 2-sphere, $d\Om^{2}_{S^{2}}$, the 
``dilaton field'' $X$ and a two dimensional metric $g_{\al \be}$ with signature 
$(+,-)$, the Hilbert-Einstein action, supplemented by an action in which the 
scalar field $S$ is coupled minimally (in $d=4$) to gravity, becomes 
equivalent to a two dimensional dilaton theory \cite{tih84} 
\eq{
L_{\text{dil}} = \frac{4\pi}{\ka} \int d^2x \sqrt{-g} \left[ X R + \frac{
\left(\nabla X\right)^2}{2X} - 2 + \ka X \left( \nabla S \right)^2 \right].
}{dil}
$R$ denotes the $2d$ curvature, $\left( \nabla X\right) ^{2}=g^{\al \be }
\partial _{\al}X\partial_{\be}X$, and $\kappa = 8\pi G_{N}$. 

The action (\ref{dil}) is locally and globally equivalent to a first order one 
\cite{iki93} depending on Cartan 1-forms $e^{\pm}$ (we 
denote light-cone indices with $\pm$) and $\om$ (the abelian gauge structure of 
the two dimensional spin connection $\om^a_{\quad b}=\eps^a_{\quad b} \om$ is
used explicitly), the dilaton field $X$, the auxiliary 
fields $X^{\pm}$ and the $s$-wave Klein-Gordon field $S$:
\eqa{
L_{\text{FO}} &=& - \frac{8\pi}{\ka} \int {\big [} X^+ (d - \om) \wedge e^- + 
X^- (d + \om) \wedge e^+ + \nonumber \\
&& + X d\wedge \omega - e^- \wedge e^+ {\cal V} - \frac{\ka}{2} X dS 
\wedge *dS {\big ]}. 
}{Q20}  
Actually, this equivalence holds for general dilaton theories. In the present
spherically reduced case the ``potential'' in (\ref{Q20}) becomes 
${\cal V} = -2 - \frac{X^+X^-}{2X}$. In the following we shall take $\kappa=1$
and drop the overall factor. Only in the final result the full 
$\kappa$-dependence will be restored.

\section{Path integral quantization of geometry}

Although in the present paper we consider only classical (tree level) 
processes, the path integral seems to be the most adequate language to derive 
the scattering amplitudes. As in other well-known examples (e.g. the 
Klein-Nishina formula for relativistic Compton scattering \cite{kln29}) this
formalism is much superior to a purely classical computation, because it
directly focuses on the physical observable, the $S$-matrix element, which 
leads to an immediate interpretation. In a 
series of papers \cite{hak94,klv99,gkv00} it has been 
shown that the path integral quantization, developed from the action 
(\ref{Q20}), allows the exact treatment of the geometric part for the choice 
of a temporal gauge of the Cartan variables
\eq{
\om_0 = 0, \hspace{0.5cm}e_0^-=1, \hspace{0.5cm}e_0^+=0. 
}{tg}
Previous work was restricted to minimally coupled scalars in (\ref{Q20}), i.e.
the dilaton factor $X$ in front of the matter action was omitted.

The Hamiltonian analysis in terms of the remaining field variables and 
associated conjugate momenta
\eq{
q_i = \left( \omega_1, e^-_1, e^+_1 \right), \hspace{0.5cm}p_i = \left( X, 
X^+, X^- \right),
}{cv}
together with the introduction of the path integral in phase space, suitably
extended by ghosts works here as in \cite{hak94,klv99,gkv00}. 
Though, as a consequence of the dilaton factor $X$ in (\ref{Q20}), the
structure functions of the constraint algebra acquire additional terms, but 
the nilpotent BRST charge also here resembles the one in Yang-Mills theories.
For details of the Hamiltonian analysis and path integral quantization we refer
to \cite{gru01}. Having integrated ghost fields and other canonical
variables, the effective action including sources for $q_i, p_i$ and $S$ 
differs only slightly from the one in \cite{klv99,gkv00}:
\eqa{
&& L = \int{\Big [} - \dot{p}_i q_i + q_1 p_2 - q_3 {\cal V} + \frac{p_1}{2} 
\left(\partial_1 S \partial_0 S - q_2 (\partial_0 S)^2 \right) 
\nonumber \\
&& \quad\quad+j_iq_i+J_ip_i+QS {\Big ]}.
}{eff2}
The generating functional for the Green functions reads
\eq{
Z\left[j, J, Q\right]=\int\left({\cal D}q\right)\left({\cal D}p\right)
\left({\cal D}S\right) \exp{(iL)}.
}{pi}

After the (exact) $q$- and $p$-integrals only the integration of scalars
remains. Thus, the usual perturbation theory is restricted to the incorporation
of matter fields. Separating terms of ${\cal O}\left( S^{2n}\right)$, $n>2$,
the Gaussian path integral of the terms up to ${\cal O}\left( S^{2}\right)$ 
yields a typical propagator contribution, apart from terms of ${\cal O}\left( 
\hbar \right)$, like a generalized Polyakov action and a contribution from the 
measure. As in ref. \cite{klv99} we concentrate on the (highly nontrivial) 
vertex ${\cal O} \left( S^{4}\right)$ in the perturbation expansion. It allows 
the calculation of scattering of $s$-wave scalars. This vertex
can be extracted formally from the final effective action. However, it contains
complicated multiple integrals. Hence, we use again the 
simple short cut introduced in \cite{klv99}, the idea of which we will outline 
briefly: It is sufficient to assume the second order combinations of the 
scalar field to be localized at a single point\footnote{%
Actually, the sources should be localized at different points, but
for the lowest order tree graphs -- which are our main goal -- this makes no
difference.} 
\eqa{
&& S_{0} := \frac{1}{2} \left(\partial_0 S\right)^2 = c_{0} \de (x-y), 
\label{local0} \\
&& S_{1} := \frac{1}{2}\left(\partial_0 S\right)\left(\partial_1 S
\right) = c_{1} \de(x-y),
}{local1} 
and to solve the classical equations of motion (EOM) following from the gauge 
fixed action (\ref{eff2}) up to linear order in the ``sources'' $c_0$ or
$c_1$. Then the solutions have to be substituted back into the interaction 
terms in (\ref{eff2}). Higher orders in $c_0, c_1$ would yield either loop 
contributions or vertices with at least 6 outer legs. We emphasize again that 
we are using perturbative methods in the matter sector only. Thus no a priori 
split into background- and fluctuation-metric occurs in our approach. 

\section{Classical EOM}

The solution of the classical EOM in the presence of matter from (\ref{eff2}) 
with vanishing sources\\
\seq{2.5cm}{
&& \partial_0 p_1 = p_2, \\
&& \partial_0 p_2 = p_1 S_0, \\
&& \partial_0 p_3 = 2 + \frac{p_2p_3}{2p_1},
}{4cm}{
&& \partial_0 q_1 = \frac{q_3p_2p_3}{2p_1^2} + S_1 - q_2 S_0, \\
&& \partial_0 q_2 = - q_1 - \frac{q_3p_3}{2p_1}, \\
&& \partial_0 q_3 = - \frac{q_3p_2}{2p_1},
}{eom}\\
to linear order in $c_0$ and $c_1$ is found easily:
\eqa{
p_1 (x) = && x_0 + (x_0 - y_0) c_0 y_0 h(x,y), \label{p1} \\
p_2 (x) = && 1 + c_0 y_0 h(x,y), \label{p2} \\
q_2 (x) = && 4 \sqrt{p_1} + \left(2c_0y_0^{3/2}-c_1y_0\right. \nonumber \\
&& \quad \left. + (c_1-6c_0y_0^{1/2})p_1 \right) h(x,y), \label{q2} \\
q_3 (x) = && \frac{1}{\sqrt{p_1}}.
}{q3}
Here $h(x,y) := \theta (y_0 - x_0) \de(x_1-y_1)$, corresponds to one of the
prescriptions introduced in \cite{gkv00} for the boundary values at $x_0 \to
\infty$. It turns out that the vertices below are {\em independent} of 
{\em any} such choice.
The matching conditions at $x_0=y_0$ follow from continuity properties: 
$p_1, q_2$ and $q_3$ are $C^0$ and $\partial_0 q_2(y_0+0)-\partial_0 q_2(y_0-0)
= \left(c_1 - q_2(y_0) c_0\right) \de(x_1-y_1)$. 
Integration constants which would produce an asymptotic (i.e. for $x_0 \to 
\infty$) Schwarzschild term and a Rindler term have been fixed to zero. Thus, 
a black hole may appear only at an intermediate stage (the ``virtual black 
hole'', see below), but should not act asymptotically. Due to the infinite 
range of gravity this is necessary for a proper $S$-matrix element, if
we want to use as asymptotic states spherical waves for the incoming and 
outgoing scalar particles.

\section{Line element}

The matter dependent solutions in our gauge (\ref{tg}) from (\ref{cv}), 
(\ref{q2}) and (\ref{q3}) define an effective line element
\eq{
(ds)^2 = 2 dr du + K(r,u) (du)^2,
}{dsef} 
if we identify\footnote{Note the somewhat unusual r\^ole of the indices 0 and 
1: $x_0$ is asymptotically proportional to $r^2$, thus our Hamiltonian 
evolution is with respect to a ``radius'' as ``time''-parameter.}
$u = 2\sqrt{2} x_1$ and $r = \sqrt{p_1(x_0)/2}$. It then appears in outgoing 
Sachs-Bondi form. The Killing norm
\eq{
\left. K (r,u) \right|_{x_0 < y_0} = \left(1 - \frac{2m}{r} - a r\right)
\left(1+ {\cal O}(c_0) \right),
}{kn}
with $m = \de(x_1-y_1)(-c_1 y_0 - 2 c_0 y_0^{3/2})/2^{7/2}$ and 
$a = \de(x_1-y_1)(c_1-6c_0y_0^{1/2})/2^{5/2}$, has two zeros located 
approximately at $r = 2m$ and $r = 1/a$ corresponding for positive $m$ and $a$
to a Schwarzschild horizon and a Rindler type one. In the asymptotic region 
the Killing norm is constant by fixing $\left. K (r,u)\right|_{x_0 > y_0} = 1$.

\section{Virtual black hole (VBH)}

As in \cite{gkv00} we turn next to the conserved quantity, which exists in all 
two dimensional generalized dilaton theories \cite{kus92}, even in 
the presence of matter \cite{kut99,grk00}. For SRG its geometric part reads
\eq{
{\cal C}^{(g)} = \frac{p_2p_3}{\sqrt{p_1}} - 4 \sqrt{p_1} 
}{conserved}
and by assumption it vanishes in the asymptotic region $x_0 > y_0$. A simple 
argument shows that ${\cal C}^{(g)}$ is discontinuous: $p_1$ and
$p_3$ are continuous, but $p_2$ jumps at $x_0 = y_0$. This phenomenon
has been called ``virtual black hole'' (VBH) in \cite{gkv00}. It is generic 
rather than an artifact of our special choice of asymptotic conditions. The
reason why we have chosen this name is simple: The geometric part of the
conserved quantity (\ref{conserved}) is essentially equivalent to the so-called
mass aspect function, which is closely related to the black hole mass 
\cite{grk00}. Moreover, inspection of the Killing norm (\ref{kn}) reveals, that
for very small Rindler acceleration $a$ the Schwarzschild horizon corresponds
to a BH with precisely that mass. This BH disappears in the asymptotic states 
(by construction), but mediates an interaction between them.

The idea that black holes must be considered in the $S$-matrix together
with elementary matter fields has been put forward some time ago 
\cite{tho96}. Our approach has allowed for the first time
to derive (rather than suppose) the existence of the black hole states in the
quantum scattering matrix. So far, we were able to perform actual
computations in the first non-trivial order only. The next order
calculations which should yield an insight into the information paradox are in
progress.

The solutions (\ref{p1}) and (\ref{p2}) establish
\eq{
\left. {\cal C}^{(g)} \right|_{x_0 < y_0} = 4 c_0 y_0^{3/2} \propto - m_{VBH}. 
}{vbh}
Thus, $c_1$ only enters the Rindler term in the Killing norm, but not 
the VBH mass (\ref{vbh}). 

\section{The $\boldsymbol{S^4}$ vertex}

All integration constants have been fixed by the arguments in the preceding 
paragraphs. The fourth order vertex of quantum field theory is extracted by 
collecting the terms quadratic in $c_0$ and $c_1$ replacing each by $S_0$ and 
$S_1$, respectively. 
\begin{figure}
\epsfxsize=7cm
\centerline{\epsfbox[70 210 540 360]{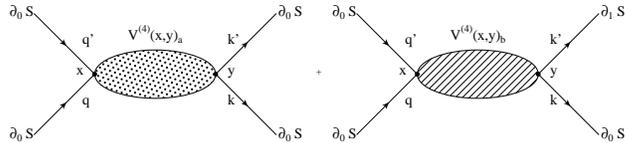}}
\caption{Total $V^{(4)}$-vertex with outer legs}
\label{fig1}
\end{figure}
\noindent The tree graphs we obtain in that way (cf. fig. \ref{fig1}) contain 
the nonlocal vertices
\eqa{
V^{(4)}_a && = \int_x\int_y S_0(x) S_0(y) \left. \left( \frac{d q_2}{d c_0} 
p_1 + q_2 \frac{d p_1}{d c_0} \right)\right|_{c_i = 0} \nonumber \\
&& = \int_x \int_y S_0(x) S_0(y) \left| \sqrt{y_0}-\sqrt{x_0} \right| 
\sqrt{x_0y_0} \nonumber \\
&& \quad \left( 3x_0+3y_0+2\sqrt{x_0y_0} \right) \de(x_1-y_1),
}{s4a}
and
\eqa{
V^{(4)}_b &&= \int_x\int_y\left. \left(S_0(y) S_1(x) \frac{d p_1}{d c_0} - 
S_0(x) S_1(y) \frac{d q_2}{d c_1} p_1 \right)\right|_{c_i = 0} 
\nonumber \\ 
&& = \int_x\int_y S_0(x) S_1(y) \left|x_0-y_0 \right| x_0 \de(x_1 - y_1), 
}{s4b}
with $\int_x := \int\limits_0^\infty dx^0\int\limits_{-\infty}^\infty dx^1$.

\section{Asymptotics}

With $t := r + u$ the scalar field satisfies asymptotically the spherical 
wave equation. For proper $s$-waves only the spherical Bessel function
\eq{
R_{k0} (r) = \frac{\sin (kr)}{kr}
}{impeig}
survives in the mode decomposition ($Dk:=4\pi k^2dk$):
\eq{
S(r,t) = \frac{1}{(2\pi)^{3/2}}\int\limits_0^{\infty} \frac{Dk}
{\sqrt{2k}} R_{k0} \left[a^+_k e^{ikt} + a^-_k e^{-ikt}\right]. 
}{asyscalmod}
With $a^\pm$ obeying the commutation relation $[a_k^-,a_{k'}^+] = \de(k-k')/
(4\pi k^2)$, they will be used to define asymptotic states 
and to build the Fock space. The normalization factor is chosen such that the 
Hamiltonian reads
\eq{
H = \frac{1}{2} \int\limits\limits_0^{\infty} Dr \left[ (\partial_t S)^2 + 
(\partial_r S)^2 \right] = \int\limits_0^{\infty} Dk a^+_k a^-_k k.
}{hamil}

\section{Scattering amplitude}

In \cite{gkv00} we had arrived at a trivial result in the massless case for 
(in $d=2$) minimally coupled scalars: Either the $S$-matrix was divergent
or -- if the VBH was ``plugged'' by suitable boundary conditions on $S$
at $r=0$ -- it vanished. Only for massive scalars we found
some finite nonvanishing scattering amplitude.

In the present physical case of $s$-waves from $d=4$ General Relativity at
a first glance it may be surprising that the simple additional factor $X$
in front of the matter Lagrangian induces fundamental changes in the 
qualitative behavior. In fact, it causes the partial differential equations 
(\ref{eom}) to become coupled, giving rise to an additional vertex 
($V^{(4)}_{b}$). 

After a long and tedious calculation (for details see \cite{gru01,fis01}) for
the $S$-matrix element with ingoing modes $q, q'$ and outgoing
ones $k, k'$
\eq{
T(q, q'; k, k') = \frac{1}{2} \left< 0 \left| a^-_ka^-_{k'} \left(V^{(4)}_a 
+ V^{(4)}_b \right) a^+_qa^+_{q'}\right| 0 \right> 
}{T}
having restored the full $\ka$-dependence we arrive at
\eq{
T(q, q'; k, k') = -\frac{i\ka\de\left(k+k'-q-q'\right)}{2(4\pi)^4
|kk'qq'|^{3/2}} E^3 \tilde{T}
}{RESULT}
with $E=q+q'$,
\eqa{
&& \tilde{T} (q, q'; k, k') := \frac{1}{E^3}{\Bigg [}\Pi \ln{\frac{\Pi^2}{E^6}}
+ \frac{1} {\Pi} \sum_{p \in \left\{k,k',q,q'\right\}}p^2 \ln{\frac{p^2}{E^2}} 
\nonumber \\
&& \quad\quad\quad\quad\quad\quad \cdot {\Bigg (}3 kk'qq'-\frac{1}{2}
\sum_{r\neq p} \sum_{s \neq r,p}\left(r^2s^2\right){\Bigg )} {\Bigg ]},
}{feynman}
and $\Pi = (k+k')(k-q)(k'-q)$. The interesting part of the scattering 
amplitude is encoded in the scale independent factor $\tilde{T}$. 

\section{Discussion}

The simplicity of (\ref{feynman}) is quite surprising, in view of the fact that
the two individual contributions (cf. figure \ref{fig1}) are not only vastly 
more complicated, but also divergent. This precise cancellation urgently asks 
for some deeper explanation.
The fact that a particular subset of graphs to a given order in perturbation
theory may be gauge dependent and even divergent, while the sum over all such
subsets should yield some finite, gauge-independent $S$-matrix is well known 
from gauge theory in particle physics (cf. e.g. \cite{kul73}).
However, it seems that only in the temporal gauge (\ref{tg}) 
one is able to integrate out the geometric degrees of freedom 
successfully. Also that gauge is free from coordinate singularities which we 
believe to be a prerequisite for a dynamical study extending across the 
horizon\footnote{%
Other gauges of this class, e.g. the Painlev{\'e}-Gullstrand gauge 
\cite{pai21} seem to be too complicated to allow an application of our 
present approach.}.

The only possible singularities occur if an outgoing momentum equals an ingoing
one (forward scattering). Near such a pole we obtain with $k=q+\eps$ and
$q\neq q'$:
\eq{
\tilde{T}(q,q';\eps) = \frac{2(qq')^2}{\eps}\ln{\left(\frac{q}{q'}\right)} + 
{\cal O}(1).
}{forward}

The nonlocality of the vertex prevents the calculation of the usual $s$-wave 
cross section. However, an analogous quantity can be defined by squaring 
(\ref{RESULT}) and dividing by the spacetime integral over the product of the 
densities of the incoming waves $\bigl( \rho=(2\pi)^{-3} \sin^2(qr)/(qr)^2 
\bigr)$: $I = \int Dr dt \rho(q)\rho(q')$, $\sigma=I^{-1}
\int_0^{\infty} DkDk' |T|^2$. Together with the introduction of 
dimensionless kinematic variables $k=E \alpha, k'=E(1-\alpha), q=E \beta, 
q'=E(1-\beta), \alpha, \beta \in [ 0,1 ]$ this yields
\begin{equation}
 \frac{d\sigma}{d\alpha}=\frac{1}{4(4\pi)^3}\frac{\kappa^2 E^2 |\tilde{T}
(\alpha, \beta)|^2}{(1-|2\beta-1|)(1-\alpha)(1-\beta)\alpha\beta}.
\label{crosssection}
\end{equation}

Our result also allows the definition of a decay rate $d^3 \Gamma/(DqDkDk')$ 
of an $s$-wave with ingoing momentum $q$ decaying (!) into three outgoing 
ones with momenta $k,k',-q'$. Clearly, lifetimes calculated in this manner
will crucially depend on assumed distributions for the momenta.

Finally, we stress that in the more general four dimensional setup of
gravitational particle scattering combinations of non-spherical modes could 
contribute to the $s$-wave matrix element. Hence, our result (\ref{RESULT}) 
does not include the full ($4d$) classical information. Nonetheless, as the 
previous discussion shows, its physical content is highly nontrivial. We
emphasize especially the decay of $s$-waves, which is a new phenomenon 
caused by the non-linearity of the underlying theory. Note that it is not 
triggered by graviton interaction, since there are no spherically symmetric 
gravitons. Still, it is caused by gravity, i.e. by gravitational self 
interaction encoded in our non local vertices. Though the existence of such 
processes may be expected on general grounds, our simple method allows us to 
calculate the corresponding amplitudes explicitly.

Our methods are useful also for other applications, such as spherically 
symmetric collapse \cite{cho93} or the polarized Gowdy model \cite{gow71}.

\section*{Acknowledgements}

This paper has been supported by projects P-12815-TPH and P-14650-TPH of the 
Austrian Science Foundation (FWF). 
One of the authors (D.V.) is grateful to the Deutsche Forschungsgemeinschaft  
(project Bo 1112/11-1) and to the Erwin Schr\"{o}dinger Institute for 
Mathematical Physics. We thank J. Wabnig for checking parts of the calculation 
of the scattering amplitude.



\end{document}